# Hidden Subsurface Oxygen Asymmetry Stabilizes Polaron Beneath the Polar $CeO_2$(100) Surface

Songda Li[1,5], Chen Zou[1,5], Liuxi Chen[1], Xiaojuan Hu[1], Ning Xu[1], Fangwen Yang[1], Zhifu Yang[1], Bingwei Chen[1], Wentao Yuan[1,3]*, Hangsheng Yang[1]*, David J. Wales[2], Sergey V. Levchenko[4], Zhong-Kang Han[1,3]*, Yong Wang[1,3]*

[1]State Key Laboratory of Silicon and Advanced Semiconductor Materials and Center of Electron Microscopy, School of Materials Science and Engineering, Zhejiang University, Hangzhou, 310027, China.

[2]Yusuf Hamied Department of Chemistry, University of Cambridge, Cambridge, CB2 1EW, UK.

[3]Institute of Fundamental and Transdisciplinary Research, Zhejiang University; Zhejiang Key Laboratory of Low-Carbon Synthesis of Value-Added Chemicals, Hangzhou, 310058, China

[4]Skolkovo Institute of Science and Technology, Bolshoy Boulevard 30/1, Moscow, Russia

[5]These authors contributed equally: Songda Li, Chen Zou.

*Corresponding authors. Email: wentao_yuan@zju.edu.cn; hsyang@zju.edu.cn; hanzk@zju.edu.cn; yongwang@zju.edu.cn

**Abstract**

Polar oxide surfaces often reconstruct to compensate their intrinsic electrostatic instability, yet identifying the atomic-scale structural motifs that govern such reconstructions and their coupling with localized electronic states remains challenging because oxygen configurations are difficult to resolve. Here, we show that a hidden asymmetric arrangement of subsurface oxygen atoms is intimately coupled with the reconstruction of the polar $CeO_2$(100) surface and the formation of subsurface small polarons. Global structure search and deep potential molecular dynamics identify a previously unknown asymmetric (1×2) reconstruction, where subsurface oxygen asymmetry creates anisotropic Ce-O coordination environments and elongated Ce-O bonds, stabilizing localized $Ce^{3+}$ small polarons beneath the surface. Enabled by oxygen-sensitive integrated differential phase contrast scanning transmission electron microscopy (iDPC-STEM), we directly visualize the predicted asymmetric subsurface oxygen configuration, while high-angle annular dark-field (HAADF) STEM resolves the accompanying asymmetric cerium arrangement and expanded interlayer spacing. These findings establish hidden subsurface oxygen asymmetry as a key structural motif coupling oxygen rearrangement, lattice reconstruction, and polaron localization, revealing a coupled oxygen-lattice-electron mechanism governing the evolution of polar oxide surfaces.

## Introduction

Metal oxides are widely utilized in catalysis,[1–4] energy conversion,[5] and electronic devices,[6] where their functionalities are intrinsically governed by the atomic structures and electronic states of their surfaces. This is because key processes, including oxygen vacancy formation,[7–11] adsorbate activation,[12,13] and charge transfer,[14,15] are governed by the interplay between surface atomic configurations and electronic structures. However, accurately determining these surface properties remains challenging due to the vast complexity of the surface configurational space. Polar oxide surfaces, which often exhibit distinct chemical and electronic properties compared with their nonpolar counterparts, are particularly challenging to study because of their intrinsic instability arising from macroscopic polarization perpendicular to the lattice plane, leading to uncompensated electrostatic potentials.[16–19] This instability triggers a variety of compensation mechanisms, among which surface reconstruction represents one of the most important pathways for reducing electrostatic energy.[20,21] Such reconstructions often involve coupled rearrangements of cations and oxygen atoms together with redistribution of localized electronic states. However, how atomic-scale oxygen configurations participate in these coupled structural and electronic transformations remains poorly understood. Within this class, the polar $CeO_2(100)$ surface is a prototypical and industrially important system in heterogeneous catalysis.[22–25] Its remarkable properties originate from the intrinsic flexibility of the Ce-O framework, including reversible oxygen storage/release, high oxygen mobility,[26,27] and the ability to accommodate localized charge carriers through $Ce^{3+}$ small polarons. Nevertheless, the surface atomic structure of $CeO_2(100)$ is highly sensitive to environmental conditions, often undergoing complex reconstructions under realistic operating environments,[28–30] which further complicates its investigation.

Extensive efforts have been devoted to understanding the structural evolution of the polar $CeO_2(100)$ surface, revealing diverse reconstructed configurations under different conditions. For example, Capdevila-Cortada *et al.* demonstrated the importance of configurational entropy in stabilizing polar $CeO_2(100)$ surfaces,[16] while Zhang *et al.* revealed a (4×6) reconstructed structure stabilized by surface electronic states.[30] These studies established that both structural disorder and electronic effects contribute to polarity compensation. Other reconstructed structures, such as $\frac{\sqrt{2}}{2}(3\times2)R45°$,[31] $(\sqrt{2}\times\sqrt{2})$,[32] (2×2),[33] and c(2×2),[34] have also been proposed using advanced surface characterization techniques, highlighting the structural diversity of polar $CeO_2(100)$ surfaces. Despite these advances, the atomic origins underlying surface reconstructions remain elusive. In particular, oxygen atoms represent a critical but experimentally inaccessible structural degree of freedom because their positions are difficult to resolve directly. Consequently, how hidden oxygen arrangements influence local coordination environments and couple with localized electronic states during surface reconstruction remains an open question.

In this work, combining global structure search, first-principles calculations, deep potential molecular dynamics, and in-situ iDPC-STEM, we uncover a previously unknown asymmetric (1×2) reconstruction on the polar $CeO_2(100)$ surface. Our calculations reveal that hidden subsurface oxygen asymmetry generates anisotropic Ce-O coordination environments and elongated Ce-O bonds, stabilizing localized $Ce^{3+}$ small polarons beneath the surface. Oxygen-sensitive iDPC-STEM subsequently resolves the predicted subsurface oxygen asymmetry, providing direct structural validation of this key feature of the theoretical model. Together, these results reveal how hidden subsurface oxygen asymmetry can govern coupled structural and electronic reconstruction at a polar oxide surface, providing a new perspective for understanding and potentially tuning redox-active surface states in ceria-based materials.

## Results and Discussion

### Machine-learning-assisted multiscale prediction of subsurface oxygen asymmetry and polaron localization

Polar $CeO_2(100)$ surfaces are highly susceptible to oxygen rearrangement under reducing environments, yet the hidden oxygen configurations that determine their reconstructed structures remain unresolved. To uncover these oxygen-driven structural motifs, we performed a global structure search using the particle swarm optimization (PSO) algorithm[35,36] (see the Methods section for details), systematically exploring different Ce/O compositions and surface configurations. By systematically varying the Ce/O ratio, we generated approximately 2,000 candidate surface structures and evaluated their stability through first-principles calculations under different oxygen chemical potentials. Among approximately 2,000 sampled structures spanning eight Ce/O compositions (Fig. 1a and Fig. S1), an oxygen-deficient $Ce_2O_3$ configuration exhibiting an ordered (1×2) reconstruction (red line) emerges as the most stable surface phase over a broad oxygen chemical potential range. The stability of this phase originates from a specific rearrangement of subsurface oxygen atoms rather than from a particular external chemical environment.

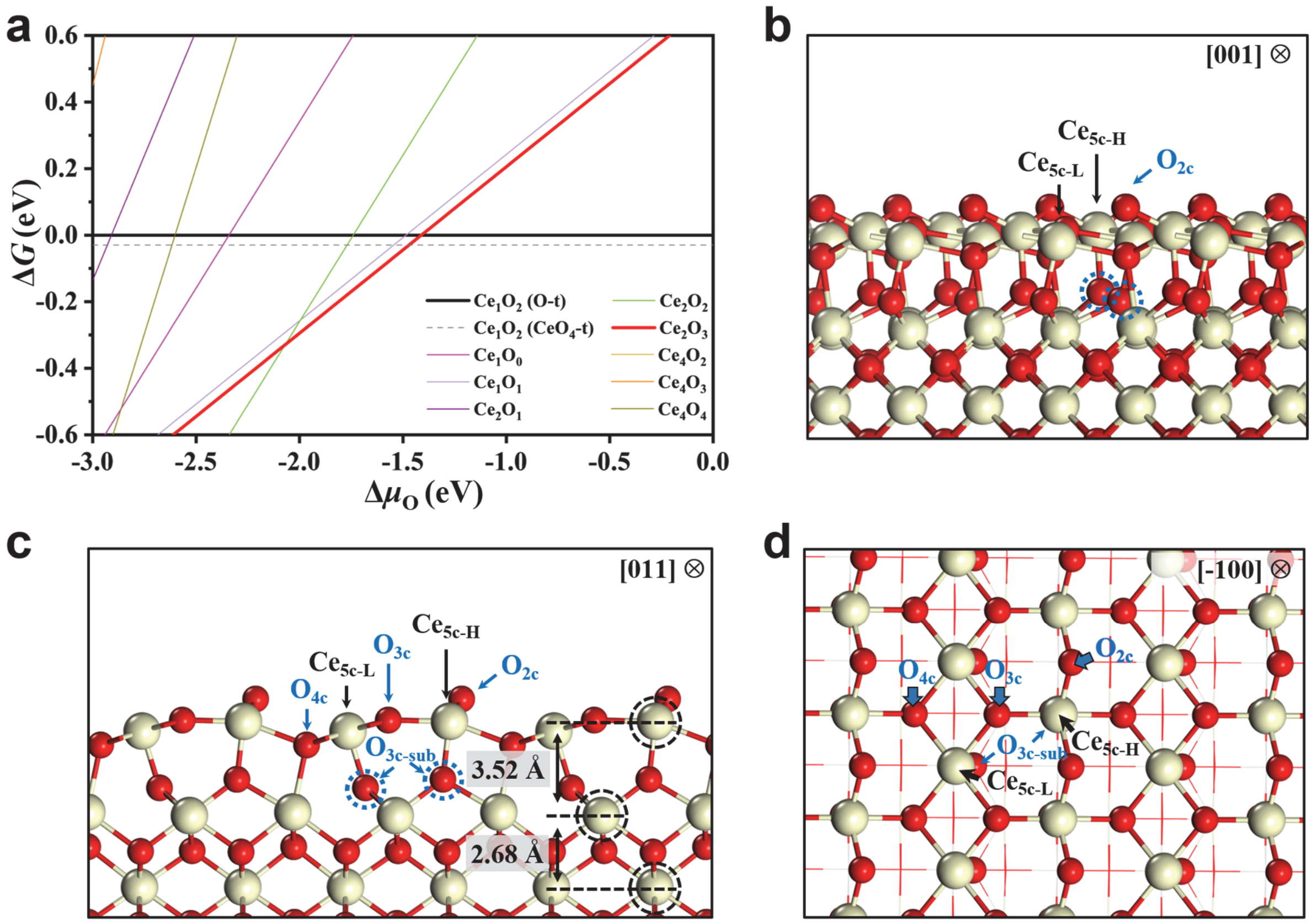

**Fig. 1 | Atomic structure of the reconstructed $CeO_2$(100)-(1×2) surface. a,** Relative surface formation energies ($\Delta G$) of structures with different compositions as a function of the oxygen chemical potential ($\Delta\mu_O$). **b-d,** Atomic structure of the most stable reconstructed surface viewed along the [001] (**b**), [011] (**c**), and [-100] (**d**) zone axes. White and red spheres represent Ce and O atoms, respectively. The two blue dashed circles in each of (**b**) and (**c**) indicate the asymmetric subsurface O sites.

The predicted reconstruction reveals a coupled surface-subsurface structural motif characterized by alternating Ce displacements and asymmetric oxygen coordination, as shown in Fig. 1b-d. Viewed along the [001] direction, the outermost Ce atoms exhibit an alternating high-low arrangement with a (1×2) periodicity. This asymmetric reconstruction originates from inequivalent oxygen environments. The higher-positioned five-coordinated Ce ($Ce_{5c\text{-}H}$) sites are coordinated by two two-coordinated surface oxygen ($O_{2c}$) atoms, two three-coordinated surface oxygen ($O_{3c}$) atoms, and one three-coordinated subsurface ($O_{3c\text{-}sub}$) atom, whereas the lower-positioned five-coordinated Ce ($Ce_{5c\text{-}L}$) sites are coordinated by two surface $O_{3c}$ atoms, two surface $O_{4c}$ atoms, and one $O_{3c\text{-}sub}$ atom (Fig. 1c and 1d). The two asymmetric $O_{3c\text{-}sub}$ sites are highlighted by blue dashed circles in Figs. 1b and 1c. The oxygen-induced structural relaxation further expands the spacing between the $Ce_{5c\text{-}H}$ surface layer and the underlying Ce layer to 3.52 Å (Fig. 1c), substantially larger than the bulk Ce layer separation of 2.68 Å. The dynamic stability of this reconstructed surface is further supported by 30-ps *ab initio* molecular dynamics (AIMD) simulations performed at temperatures ranging from 300 to 900 K (Fig. S2). Importantly, this reconstruction establishes an oxygen-coordination asymmetry within the subsurface lattice, which provides the structural origin for the selective localization of $Ce^{3+}$ polarons associated with oxygen deficiency.

To determine how the hidden oxygen asymmetry governs polaron localization, we initialized occupied Ce 4*f* states at different surface and subsurface sites and fully relaxed four representative configurations with the same surface composition and periodicity (Fig. 2). The corresponding subsurface O pair, highlighted by blue dashed circles, exhibits distinct local relaxations among these polaron configurations. The ground state contains one five-coordinate surface $Ce^{3+}$ and one six-coordinate subsurface $Ce^{3+}$, revealing a cooperative surface-subsurface polaron arrangement enabled by the reconstructed oxygen framework. Around the subsurface polaron, the surrounding oxygen framework undergoes a cooperative asymmetric relaxation, generating two inequivalent oxygen environments within the reconstructed lattice. This oxygen rearrangement produces anisotropic Ce-O coordination environments and local lattice distortions, which are coupled with the stabilization of the subsurface $Ce^{3+}$ polaron.

Further comparison of the competing polaron configurations reveals that the ground-state preference cannot be attributed to a single structural descriptor, such as the oxygen arrangement or the coordination number of the Ce sites. When both polarons are localized at surface Ce sites (Fig. 2b), the subsurface oxygen configuration differs from that of the ground state, leading to an increased energy of 0.11 eV. The higher-energy configurations in Figs. 2c and 2d also preserve asymmetric oxygen arrangements, but their subsurface $Ce^{3+}$ centers experience distinct local coordination environments and lattice distortions, resulting in energy penalties of 0.14 and 0.54 eV, respectively. These comparisons indicate that the energetic ordering is not controlled by oxygen asymmetry alone, but instead emerges from the cooperative optimization of the hidden subsurface oxygen arrangement, the surrounding Ce-O coordination network, and the associated local lattice relaxation. Such oxygen-lattice coupling generates the specific subsurface environment required for stabilizing the ground-state vacancy-induced $Ce^{3+}$ polaron.

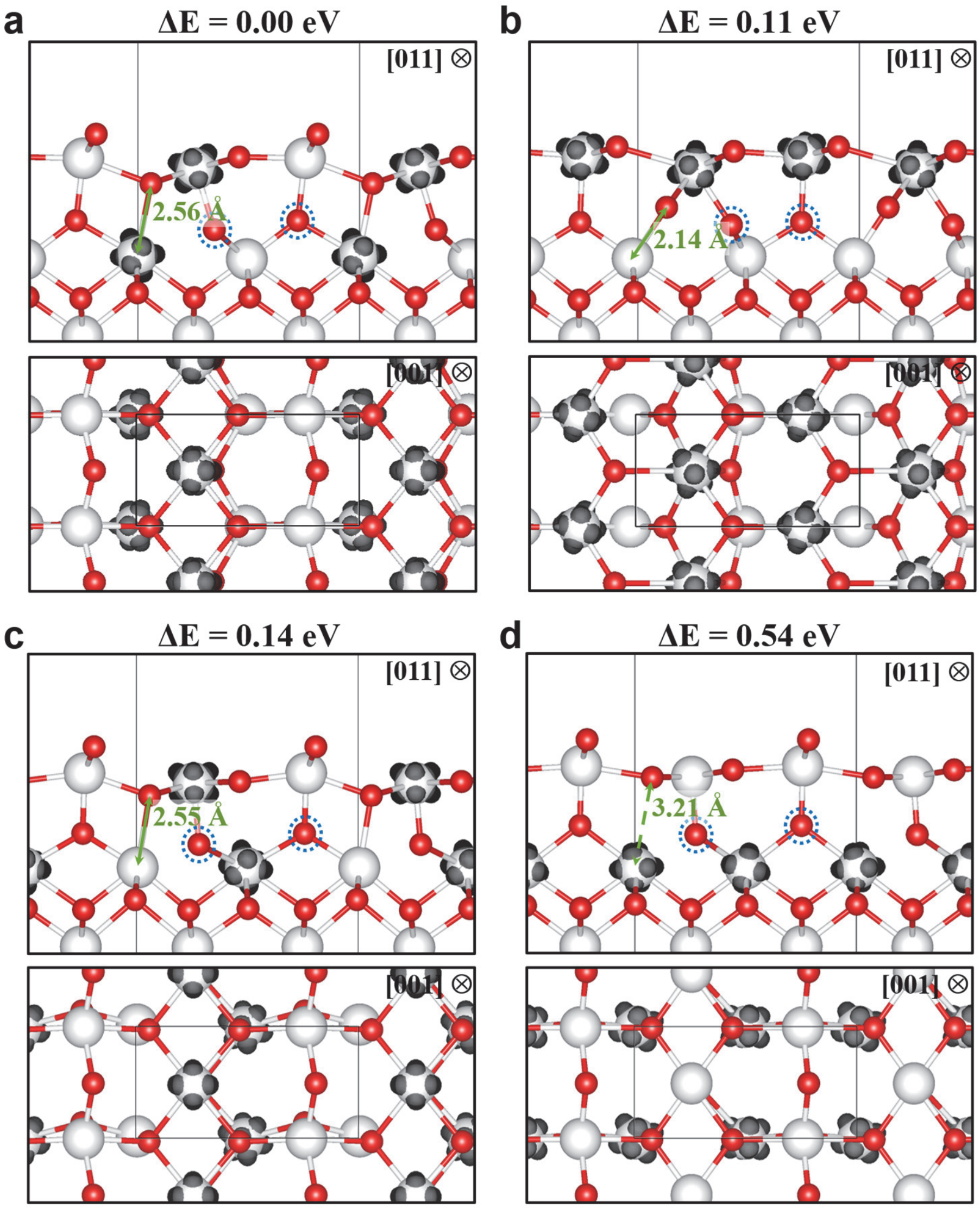

**Fig. 2 | Atomic structures and relative energies of polaron configurations on the reconstructed $CeO_2$(100)-(1×2) surface. a-d**, Side views along the [011] zone axis and corresponding top views of four fully relaxed polaron configurations with the same surface composition and (1×2) periodicity, arranged according to their relative energies. Black isosurfaces represent the spin density associated with occupied Ce 4*f* states. White and red spheres represent Ce and O atoms, respectively. Green arrows indicate selected Ce-O distances. Blue dashed circles indicate the subsurface O sites.

The thermodynamic stability identified from structure searching does not necessarily guarantee that a reconstructed structure can emerge under realistic thermal conditions. We therefore examined the structural evolution of the predicted $CeO_2$(100)-(1×2) reconstruction using deep potential molecular dynamics (DPMD) simulations.[37,38] The deep potential model was trained on 139,307 data points obtained from DFT calculations and AIMD trajectories. The resulting model exhibited a root-mean-square error (RMSE) of approximately 3 meV per atom in energy predictions (Figs. S3 and S4), demonstrating sufficient accuracy for describing the structural evolution of the $CeO_2$(100) surface. DPMD simulations were subsequently performed at temperatures ranging from 300 to 900 K using various initial configurations with the same stoichiometry as the predicted $CeO_2$(100)-(1×2) reconstruction.

Starting from a pristine $CeO_2$(100) surface with a single surface oxygen atom manually removed as the initial configuration, we performed a 500-ps DPMD simulation at 900 K. As shown in Fig. 3a, the surface underwent rapid cooperative rearrangement of oxygen and cerium atoms, accompanied by redistribution of the oxygen vacancy and emergence of the ordered $CeO_2$(100)-(1×2) reconstruction within 1.25 ps. Concurrently, the subsurface O pair, highlighted by blue dashed circles, evolved from symmetry-equivalent positions in the initial Structure I to asymmetric positions in Structure V. The reconstructed surface remained structurally persistent throughout the 500-ps simulation, despite transient local fluctuations. These results indicate that the predicted reconstruction can spontaneously emerge under thermal fluctuations rather than representing only a static minimum obtained from structure searching. The robustness of this structural evolution was further examined using disordered atomic models generated with the same surface composition. At 300 K (Fig. 3b), the initially disordered surface, characterized by oxygen dangling bonds at 116.33 ps, gradually reorganized into the (1×2) reconstruction by 508.32 ps and maintained its structural integrity throughout the rest of the 1000-ps simulation. At 900 K (Fig. 3c), enhanced atomic fluctuations accelerated the structural reorganization, with the (1×2) reconstruction emerging within 1.62 ps after substantial rearrangement (Structure II), followed by persistent structural stability during the simulation. These simulations demonstrate that the $CeO_2$(100)-(1×2) reconstruction is not only thermodynamically favorable but also thermally accessible and dynamically stable across different initial configurations, including both a reduced surface containing an isolated oxygen vacancy and randomly generated disordered surfaces. The spontaneous emergence of the same reconstructed motif highlights the intrinsic coupling between oxygen rearrangement and lattice relaxation under realistic thermal conditions.

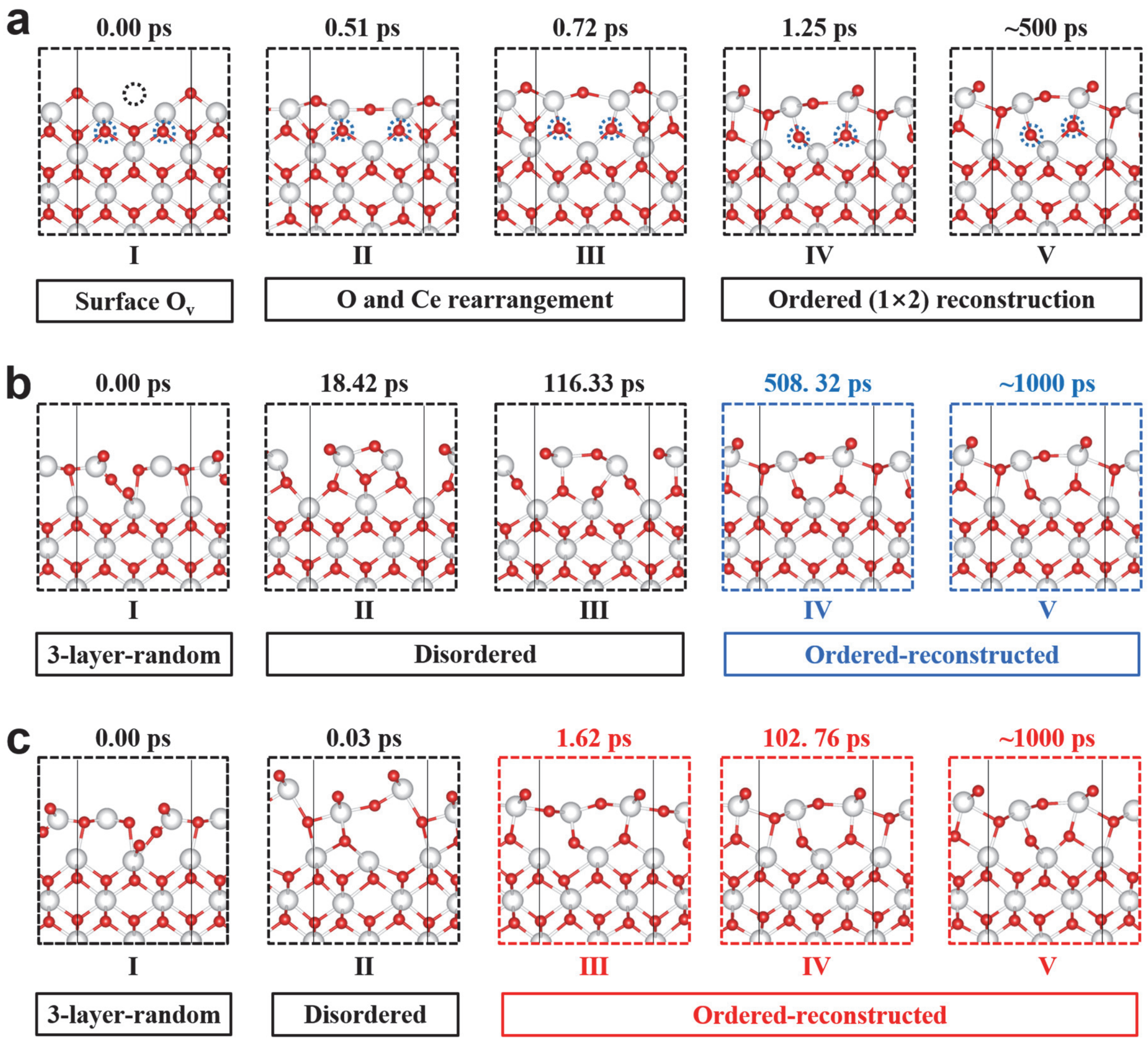

**Fig. 3 | Structural evolution from the deep potential molecular dynamics (DPMD) simulations for different initial models with the same composition as the ordered $CeO_2$(100)-(1×2) surface reconstruction, viewed along the [011] zone axis. a,** A 500-ps simulation at 900 K for the pristine $CeO_2$(100) surface containing a single surface oxygen vacancy. The black dashed circle marks the initial surface oxygen vacancy. Blue dashed circles mark the pair of subsurface O sites that evolve from symmetry-equivalent positions in structure I to asymmetric positions in the reconstructed structure V. **b, c,** 1000-ps simulations at 300 K (**b**) and 900 K (**c**) for randomly generated disordered surfaces, respectively.

## Direct in-situ iDPC-STEM imaging of subsurface oxygen asymmetry

The theoretically predicted $CeO_2$(100)-(1×2) reconstruction was directly examined using in-situ atomic-resolution STEM. Because the key structural variable identified by theory is the hidden subsurface oxygen arrangement, complementary HAADF STEM and oxygen-sensitive iDPC-STEM imaging were employed to resolve the cerium framework and oxygen sublattice, respectively. As shown in Figs. 4a-d, the experiments were conducted using a spherical aberration-corrected STEM (FEI Spectra 300, 300 kV), equipped with a double-tilt heating holder (Wildfire D6, DENS Solutions), under a low beam current of 2 pA. The column was maintained in a high vacuum (~$10^{-5}$ Pa), without significant change during the entire STEM experiment. To avoid the influence of surface adsorbates and other contaminants, before in-situ observation, the samples were subjected to a series of in-situ treatments such as electron beam shower and heating pre-treatment (see Methods for experimental details). The $CeO_2$ samples were in-situ annealed at 1173 K in vacuum for at least 10 minutes. At room temperature (293 K), the $CeO_2$ nanocubes predominantly exhibit bulk-truncated (100) surfaces with an interplanar spacing of 0.27 nm (Figs. 4a, 4c, and S5), consistent with previous observations[39]. Following annealing at 1173 K, the $CeO_2$(100) surface develops a reconstructed structure with an expanded interlayer spacing of 0.35 nm between the outermost and subsurface Ce layers, as shown in Figs. 4b and 4d, in excellent agreement with the theoretical prediction. The preservation of the bulk-like 2.7 Å spacing between deeper Ce layers confirms that the reconstruction is confined to the near-surface region.

Notably, the alternating high-low configuration of Ce atoms in the outermost layer, observed experimentally, matches the predicted atomic structure. This agreement confirms the predicted asymmetric cerium displacement associated with the reconstructed lattice. This result is further supported by intensity line profiles in Fig. 4e, where the peak positions in the yellow curve align with valleys in the purple curve, indicating lateral alternation in Ce atom height without positional overlap. The simulated high-resolution HAADF STEM image (right of Fig. 4b) reproduces the experimental HAADF STEM contrast well, further validating the $CeO_2$(100)-(1×2) reconstructed surface model. The Bright-field (BF) STEM image in Fig. S6 shows a clean and well-ordered surface reconstruction. Additional experimental evidence from another $CeO_2$ nanocube, imaged along the [011] direction (Figs. 4d and 4f), confirms the same surface reconstruction, reinforcing the reliability of the predicted $CeO_2$(100)-(1×2) structural model.

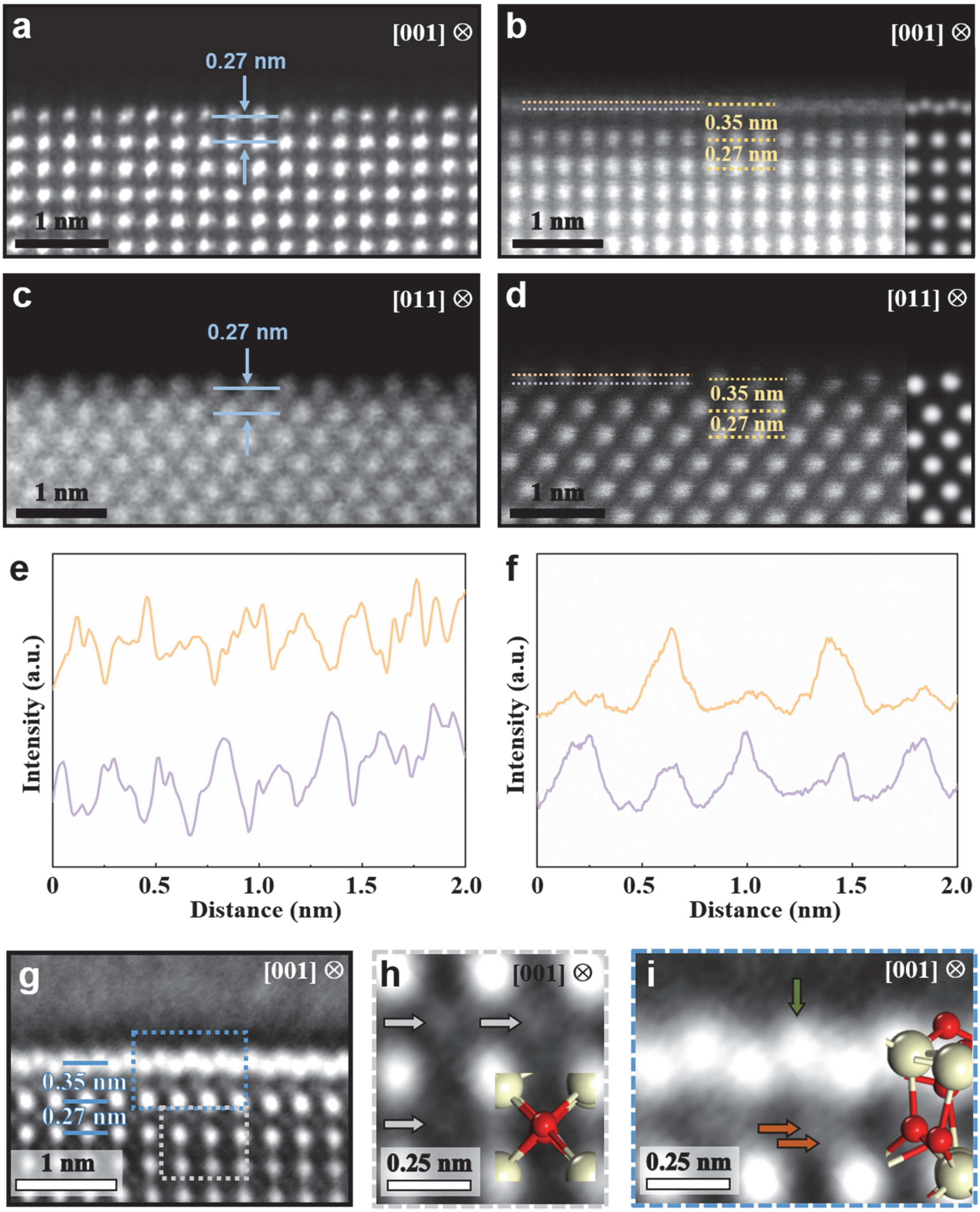

**Fig. 4 | In-situ STEM characterization of the reconstructed $CeO_2$(100)-(1×2) surface. a, c,** Atomic-resolution HAADF (high-angle annular dark-field) STEM (scanning transmission electron microscopy) images of $CeO_2$(100) surface along [001] (**a**) and [011] (**c**) zone axes at 293 K. **b, d,** Comparison of experimentally observed (left) and theoretically simulated (right) HAADF STEM images of the reconstructed $CeO_2$(100) surface along the [001] (**b**) and [011] (**d**) zone axes at 1173

K. **e, f,** Corresponding intensity line profiles extracted from the regions marked by colored dashed lines in (**b**) and (**d**), respectively. **g,** Atomic-resolution iDPC-STEM (integrated differential phase contrast scanning transmission electron microscopy) image of the reconstructed $CeO_2$ (100) surface along the [001] zone axis at 1173 K. **h, i,** Magnified areas of the iDPC-STEM images shown in (**g**), corresponding to the near-bulk region (**h**) and the surface region (**i**), respectively. The insets in (**h**) and (**i**) show the projected atomic models of the bulk-like $CeO_2$ lattice and the reconstructed $CeO_2$(100)-(1×2) surface, respectively. White and red spheres represent Ce and O atoms, respectively. White arrows in (**h**) mark the weaker-contrast bulk-like O columns. Green and orange arrows in (**i**) indicate the stronger-contrast higher-lying surface Ce columns and the weaker-contrast inequivalent subsurface O columns, respectively.

Beyond the cerium framework resolved by HAADF STEM, oxygen-sensitive iDPC-STEM provides direct access to the hidden oxygen sublattice. Recent advances in iDPC-STEM have substantially improved the visibility of light-element sublattices alongside heavier atomic columns, enabling atomic-scale imaging of oxygen-containing frameworks.[40,41] As shown in Figs. 4g-i, the iDPC-STEM images reveal oxygen contrast within the reconstructed surface region, with Fig. 4g providing an overview of the reconstructed surface and the near-bulk region used for subsequent analysis. The assignment of the oxygen-related contrast is first validated in the near-bulk region (Fig. 4h), where the stronger contrast maxima coincide with the projected Ce columns of the fluorite $CeO_2$ lattice, whereas the weaker contrast features located between four neighboring Ce columns closely match the expected projected positions of the bulk O columns. This correspondence identifies the weaker iDPC features as signatures of the oxygen sublattice. In the reconstructed surface region (Fig. 4i), these oxygen-related features deviate markedly from their near-bulk arrangement, allowing direct visualization of the asymmetric subsurface oxygen arrangement predicted by theory. The experimentally observed oxygen configuration agrees with the calculated structure, including the inequivalent oxygen positions associated with the reconstructed (1×2) periodicity. This direct observation provides direct experimental evidence supporting subsurface oxygen asymmetry as the structural origin underlying the coupled lattice reconstruction and polaron localization.

**Electronic descriptors of asymmetric reconstruction stability**

The stabilization of subsurface polarons by the reconstructed oxygen framework implies a distinct electronic reorganization within the Ce-O lattice. We therefore examined the electronic structure of the $CeO_2$(100)-(1×2) reconstruction to identify the electronic origin associated with polaron and reconstruction stabilization. The projected density of states (PDOS) reveals characteristic localized $Ce^{3+}$ states in the reconstructed surface. The occupied Ce 4*f* states appear as distinct in-gap features below the Fermi level, confirming the formation of localized small polarons. Notably, these occupied Ce 4*f* states overlap with O 2*p*-derived states, generating hybridized occupied states near the valence-band edge. Such Ce 4*f*-O 2*p* coupling reflects the strong interaction between localized electrons and the reconstructed oxygen lattice, providing an electronic signature of the oxygen-lattice-electron coupling identified above. The individual PDOS contributions from representative $Ce^{3+}$ sites (purple and green curves in Fig. 5a) align with the main features in the total Ce 4*f* states.

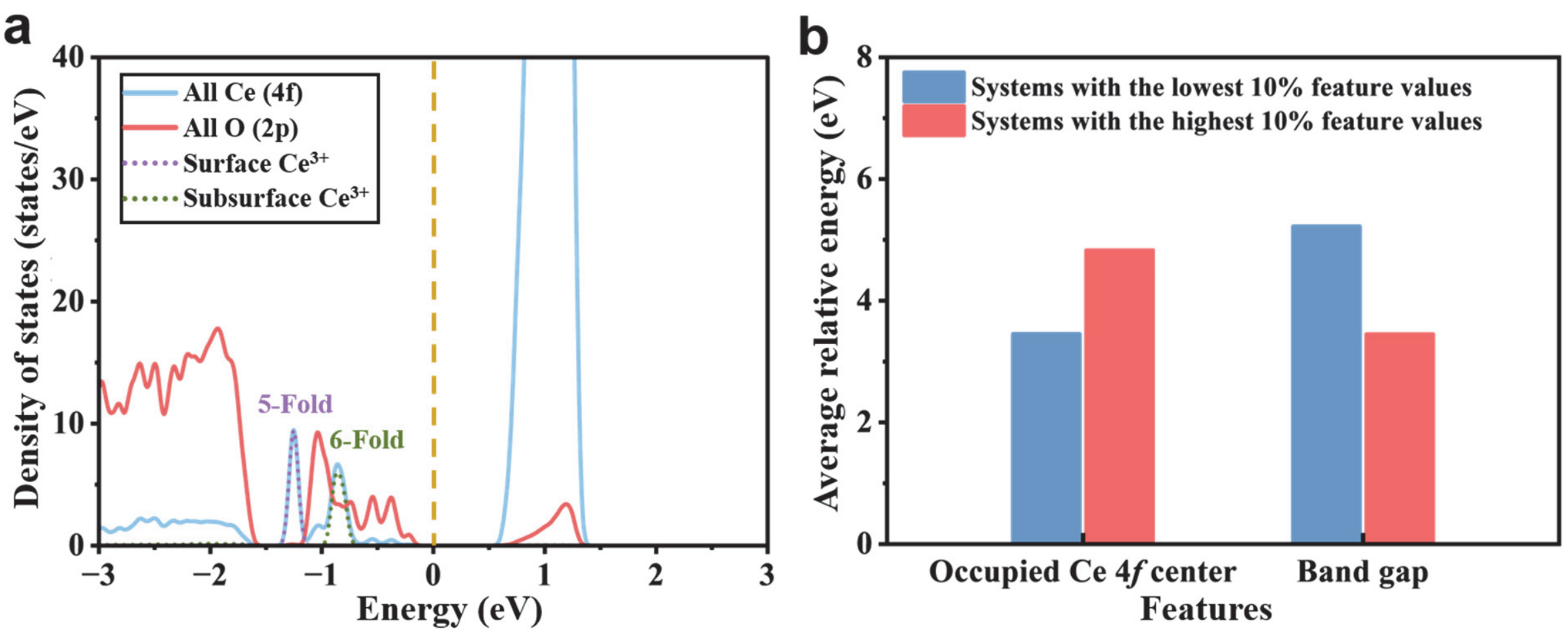

**Fig. 5 | Electronic signatures associated with the asymmetric $CeO_2$(100)-(1×2) reconstruction. a,** Projected density of states (PDOS) of O 2*p* and Ce 4*f* states, together with site-resolved $Ce^{3+}$ 4*f* contributions from representative $Ce^{3+}$ polaronic sites. **b,** Relative energies of reconstructed configurations categorized according to the occupied Ce 4*f* center and occupied O 2*p* to unoccupied Ce 4*f* band gap.

Beyond the ground-state structure, we further analyzed 100 reconstructed configurations with identical compositions to identify electronic factors governing their relative stability. By comparing the lowest 10% and highest 10% configurations according to their occupied Ce 4*f* center, we found that structures with deeper occupied Ce 4*f* states exhibit lower relative energies (Fig. 5b), indicating that enhanced localization of $Ce^{3+}$ polaronic states is associated with increased structural stability. A deeper Ce 4*f* center reflects stronger stabilization of excess electrons at reduced Ce sites and a more favorable local electronic environment. Similarly, configurations within the highest 10% of the occupied O 2*p*-unoccupied Ce 4*f* band gap show lower relative energies than those within the lowest 10%, suggesting that larger O 2*p*-Ce 4*f* band gap is associated with enhanced stabilization of the reconstructed surface. This trend suggests that an increased band gap between occupied oxygen 2*p* states and unoccupied Ce 4*f* states suppresses unfavorable electron excitation and reduces electronic instability within the Ce-O framework. These results reveal that the stability of the surface reconstructions is governed by the cooperative optimization of localized Ce 4*f* polaron states and the band gap between oxygen 2*p* states and Ce 4*f* states.

## Conclusion

In summary, we uncover a previously unknown asymmetric (1×2) reconstruction of the polar $CeO_2$(100) surface and identify hidden subsurface oxygen asymmetry as the key structural motif governing its formation. By combining global structure search, first-principles calculations, deep potential molecular dynamics, and in-situ atomic-resolution STEM, we demonstrate that subsurface oxygen rearrangement creates anisotropic Ce-O coordination environments and lattice distortions that stabilize localized $Ce^{3+}$ small polarons beneath the surface. Oxygen-sensitive iDPC-STEM directly resolves the predicted subsurface oxygen configuration, while HAADF STEM confirms the associated cerium displacement and surface lattice expansion. Electronic structure analysis further reveals that the stability of the surface reconstructions arises from the cooperative optimization of localized Ce 4*f* polaron states and O 2*p*-Ce 4*f* band gap. These findings establish hidden oxygen asymmetry as an essential structural degree of freedom linking oxygen rearrangement, lattice reconstruction, and electronic localization, providing a framework for understanding the coupled oxygen-lattice-electron evolution of polar oxide surfaces.

## Methods

### Density functional theory (DFT) calculations and global structure search

DFT calculations were performed using the Vienna Ab initio Simulation Package (VASP) code.[42,43] The generalized gradient approximation (GGA) of the Perdew-Burke-Ernzerhof (PBE) functional[44] and the projector augmented-wave (PAW) method[45] were employed. A plane-wave basis set with a cut-off energy of 400 eV was used. The convergence criterion for the self-consistent field calculations was $10^{-5}$ eV. All structures were relaxed until the maximum force on the atom was lower than 0.05 eV $Å^{-1}$. Grids of (4×2×1) and (8×4×1) k-points were used for the structural optimization and Density of States (DOS) calculations. Spin polarization was considered in all calculations and the thickness of the vacuum layer for all models was set as 15 Å.

All calculations for these $CeO_2$ models in the present study were performed by the DFT+U method. The Hubbard term U = 5.0 eV was added to diminish the self-interaction error and to properly localize the Ce 4*f* states[46,47]. Periodic slabs with (1×2) surface unit cells were applied to build the $CeO_2$(100) surface model. The bottom two layers of the model of the $CeO_2$(100) surface were frozen to their bulk-like positions during geometry optimizations. Half of the oxygen atoms were removed from the outermost plane of $CeO_2$(100), which then became stoichiometric.

The global structural searching was performed using the particle swarm optimization (PSO) algorithm implemented in the Calypso code[35,36], which involved placing different numbers of additional Ce and O atoms within a 9 Å height range on the optimized $CeO_2$(100) surface (Fig. S1). The resulting structures were then optimized using DFT while maintaining the same fixed number of layers. Systematically, each Ce/O ratio was investigated for approximately 250 structures, with 25 generations per composition, yielding over 2000 fully relaxed structures. Additionally, to control the ratio of random structures generated by the PSO algorithm, the PSO ratio parameter was set to 0.6.

The relative surface energy ($\Delta G$) is used as a criterion to evaluate the stability of a surface structure.[48,49]

$$\Delta G = \frac{1}{A}\left[G_{\text{total}} - G_{\text{slab}} - i\, G_{\text{bulk}} + (2i - j)\mu_{\text{O}}\right] \quad (1)$$

where $G_{\text{total}}$, $G_{\text{slab}}$, and $G_{\text{bulk}}$ are the Gibbs free energy of the searched surface structure, the unreconstructed surface structure, and the energy per bulk $CeO_2$ unit, respectively. *i* and *j* represent the difference in cerium and oxygen atom numbers between the configuration located in the search and the unreconstructed surface structure, respectively. Accordingly, the Ce/O compositional labels used in Fig. 1a denote the relative compositions of the searched structures with respect to the stoichiometric O-t reference surface ($Ce_{10}O_{20}$). $A$ represents the supercell size of standard surface slab from $CeO_2$ bulk. The surface formation energy values can be approximated by the corresponding internal energies[50], so Eq. (1) can be written as:

$$\Delta G = \frac{1}{A}\left[E_{\text{total}} - E_{\text{slab}} - i\,E_{\text{bulk}} + (2i - j)\mu_{\text{O}}\right], \tag{2}$$

where $\mu_{\text{O}}$ is the chemical potential of oxygen[51], which can be calculated as:

$$\mu_{\text{O}} = \frac{1}{2}\left[E_{\text{O}_2} + \Delta H_{\text{O}_2}(T,P^0) - T\Delta S_{\text{O}_2}(T,P^0) + k_{\text{B}}T\ln\left(\frac{P}{P^0}\right)\right] = \frac{1}{2}E_{\text{O}_2} + \Delta\mu_{\text{O}}, \tag{3}$$

where $E_{\text{O}_2}$ is the internal energy of gaseous oxygen. $k_{\text{B}}$, $P$, $P^0$, and $T$ are the Boltzmann constant, oxygen partial pressure, standard atmospheric pressure, and temperature, respectively. $\Delta H_{\text{O}_2}(T,P^0)$ and $\Delta S_{\text{O}_2}(T,P^0)$ are the enthalpy and entropy changes of gaseous oxygen, which can be found in the database. By comparing the calculated $\Delta G$ for structures with different compositions, their relative stability can be obtained.

**Ab initio molecular dynamics (AIMD) simulations**

AIMD simulations were initiated from optimized configurations and run for more than 30 ps with a 1 fs timestep. The simulations were performed in the canonical (NVT) ensemble, applying Nosé-Hoover thermostats with temperatures set as 300 K, 600 K, and 900 K.[52,53]

**Deep potential molecular dynamics (DPMD)**

The smooth version of the DeePMD-kit package[37,38] was used to construct the deep potential models. The primary data sets for the training were generated from AIMD and PSO searching based on DFT calculations. DP-GEN was also used to refine the quality of the model.[54] Four deep-potential (DP) models were trained using DeePMD-kit. The models share the same training data set but differ in the random seed used to initialize the parameters. The cut-off radius was set to 7.0 Å. The inverse distance $1/r$ started to decay smoothly from 0.5 Å. The size of the embedding net was (25 50 100), and the size of the fitting net was (240 240 240). The training steps were set to 2,000,000. In each iteration, three steps, exploration, labeling, and training, were carried out consecutively. The maximum standard deviation of force $\epsilon$ predicted by the four DP models was used to select the new data for active learning, the structures generated by DPMD simulations that satisfied the condition 0.15 eV/Å $\leq \epsilon <$ 0.35 eV/Å were chosen as candidate structures. Our final data set to train the model contained 139,307 data points.

The root-mean-square error (RMSE) was used to validate the accuracy of the models. At the end of the training, the RMSEs for the energies were only around 3 meV/atom, and the RMSEs for the forces were around 64 meV/Å. Correlation plots for the comparison of energies and forces predicted by DPMD and those computed from DFT calculations are shown in Fig. S3. The model was also checked by comparing the calculated radial distribution functions (RDF) from AIMD and DPMD simulations for the O-Ce pair, which agree well, indicating that our model was able to represent the local atomic interactions with DFT accuracy (Fig. S4).

**Synthesis of $CeO_2$ nanocubes**

To validate the theoretically predicted reconstructed structure of the $CeO_2$(100) surface, a $CeO_2$ nanocube, which mainly exposes (100) surfaces, was synthesized through a hydrothermal method.[39] Specifically, 0.868 g $Ce(NO_3)_3 \cdot 6H_2O$ and 9.6 g sodium hydroxide were fully dissolved in 5 mL and 35 mL of deionized water, respectively. The $Ce(NO_3)_3$ solution was dropped into the NaOH solution. The mixed solution was stirred continuously at room temperature for 30 minutes and transferred to a 50 mL Teflon-lined stainless-steel autoclave and hydrothermally treated at 453 K for 24 hours. After the hydrothermal treatment, the centrifugal separation was employed and the precipitate was then washed several times with deionized water and ethanol centrifugation. Finally, the precipitates were dried overnight in an air environment at 333 K.

**In-situ scanning transmission electron microscopy (STEM)**

Ex-situ STEM characterization was performed in an FEI Titan $G^2$ 80-200 scanning transmission electron microscope (200 kV), with a spherical aberration corrector. In-situ STEM experiments (including iDPC-STEM characterization) were performed on an FEI Spectra 300 (S)TEM microscope (300 kV) and a Hitachi HF5000 environmental scanning transmission electron microscope (200 kV), both of which are equipped with an aberration corrector. As for experiments in the FEI Spectra 300 (S)TEM microscope, the convergence semi-angle was 21 mrad. The beam current was about 2 pA. During the in-situ heating characterization, a double tilt heating holder (Wildfire D6, DENS solutions) and the $SiN_x$ heating chips were used, with a heating rate of ~2 K/s. As for experiments in the Hitachi HF5000 microscope, a double tilt heating holder (Wildfire D6, DENS solutions for heating) and $SiN_x$ heating chips were used, with heating rates of ~2 K/s and ~1 K/s, respectively.

During all STEM characterizations, the $CeO_2$ nanocubes were dispersed into ethanol and dropped on the copper grids or the heating chips. For in-situ STEM characterizations, the sample was loaded onto the heating holder after synthesis. Before the in-situ STEM characterization, the sample was first subjected to a 20-minute electron beam shower to remove the majority of surface adsorbates. The electron beam current of the beam shower was 1260 pA, with an irradiation area of approximately 76 μm × 76 μm. Subsequently, the sample alone underwent a more specific in-situ pre-treatment to remove trace environmental contaminants, such as residual carbon and water molecules. This pre-treatment procedure involved annealing at 573 K for 1 hour in a pure oxygen environment (~1 Pa at the sample location), followed by cooling to room temperature and pumping under high vacuum ($10^{-5}$ Pa).

To avoid possible damage from the electron beam (e-beam) irradiation, the e-beam dose was strictly controlled below 100 e/A[2,55,56]. In addition, the target surface area was only illuminated when imaging. Regarding the vacuum quality during STEM handling, the high vacuum of $10^{-5}$ Pa was continuously maintained during the entire STEM experiment.

To avoid potential background interference from the $SiN_x$ heating chip during STEM characterizations, all key observations were exclusively performed on the edges of $CeO_2$ nanocubes suspended over the holes of the $SiN_x$ chip. These suspended regions are free of $SiN_x$ film contact, ensuring that the collected signals solely originate from the $CeO_2$ surface and adjacent vacuum layer without any support-induced artifacts.

## Data availability

All data are available in the main text or the supplementary materials.

## Acknowledgments

This work was supported by the National Key Research and Development Program of China (2023YFA1506904), the National Natural Science Foundation of China (52025011, 52422311), the Zhejiang Provincial Natural Science Foundation of China (LR23B030004), the Leading Innovative and Entrepreneur Team Introduction Program of Zhejiang (2023R01007), China

Postdoctoral Science Foundation (2025M770064), and the Fundamental Research Funds for the Central Universities. The authors thank Dr. Zhemin Wu and Dr. Ruiyang You in the Center of Electron Microscopy at Zhejiang University for TEM data analysis.

**Author contributions**

Y.W. initiated the project. Y.W., Z.H., H.Y., and W.Y. supervised this study. S.L. carried out the calculations, analyzed the data, and drafted the manuscript. C.Z. performed the in-situ STEM observations. L.C., X. H. and N. X. assisted with the calculations. F.Y., Z.Y. and B.C. helped with the experiments. W.Y., H.Y., D.J.W., and S. V. L. participated in discussion. All the authors commented on the manuscript and approved the final version. S.L. and C.Z. contributed equally to this work.

**Competing interests**

The authors declare no competing interests.

**Additional information**

**Supplementary Information** is available for this paper.

**Correspondence and requests for materials** should be addressed to Wentao Yuan, Hangsheng Yang, Zhong-Kang Han, and Yong Wang.

Supplementary Information for

# Hidden Subsurface Oxygen Asymmetry Stabilizes Polaron Beneath the Polar $CeO_2$(100) Surface

Songda Li[1,5], Chen Zou[1,5], Liuxi Chen[1], Xiaojuan Hu[1], Ning Xu[1], Fangwen Yang[1], Zhifu Yang[1], Bingwei Chen[1], Wentao Yuan[1,3]*, Hangsheng Yang[1]*, David J. Wales[2], Sergey V. Levchenko[4], Zhong-Kang Han[1,3]*, Yong Wang[1,3]*

[1]State Key Laboratory of Silicon and Advanced Semiconductor Materials and Center of Electron Microscopy, School of Materials Science and Engineering, Zhejiang University, Hangzhou, 310027, China.

[2]Yusuf Hamied Department of Chemistry, University of Cambridge, Cambridge, CB2 1EW, UK.

[3]Institute of Fundamental and Transdisciplinary Research, Zhejiang University; Zhejiang Key Laboratory of Low-Carbon Synthesis of Value-Added Chemicals, Hangzhou, 310058, China

[4]Skolkovo Institute of Science and Technology, Bolshoy Boulevard 30/1, Moscow, Russia

[5]These authors contributed equally: Songda Li, Chen Zou.

*Corresponding author. Email: wentao_yuan@zju.edu.cn; hsyang@zju.edu.cn; hanzk@zju.edu.cn; yongwang@zju.edu.cn

**The PDF file includes:**

Supplementary Figs. 1 to 6

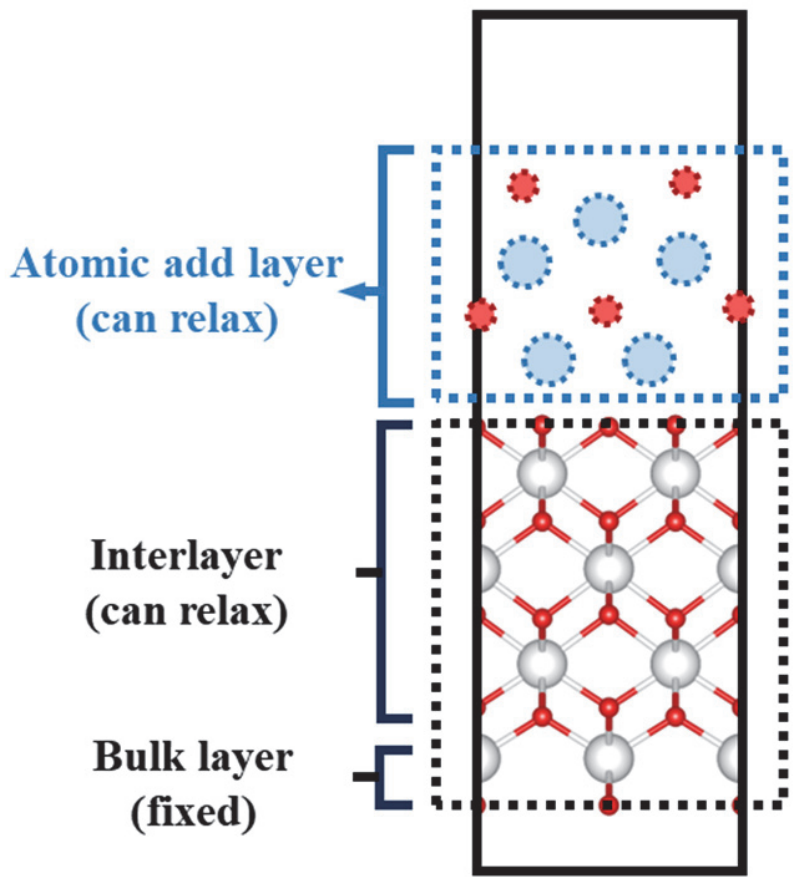

**Supplementary Fig. 1.** Setup details of initial structures for PSO and DFT calculations. Blue and White, Ce; Red, O.

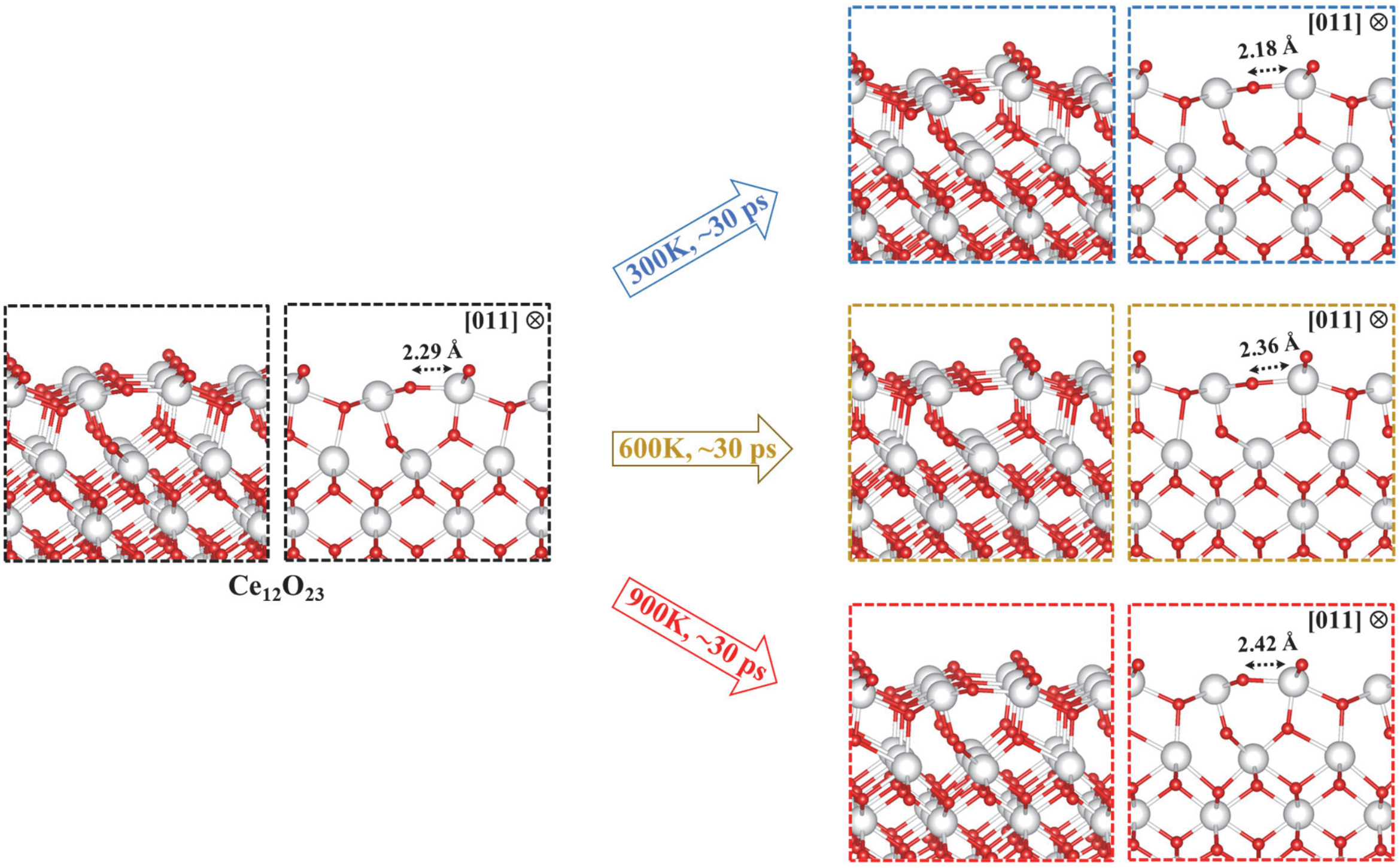

**Supplementary Fig. 2.** The stability validation results of the $CeO_2$(100)-(1×2) reconstruction by AIMD simulations at different temperatures. White, Ce; Red, O.

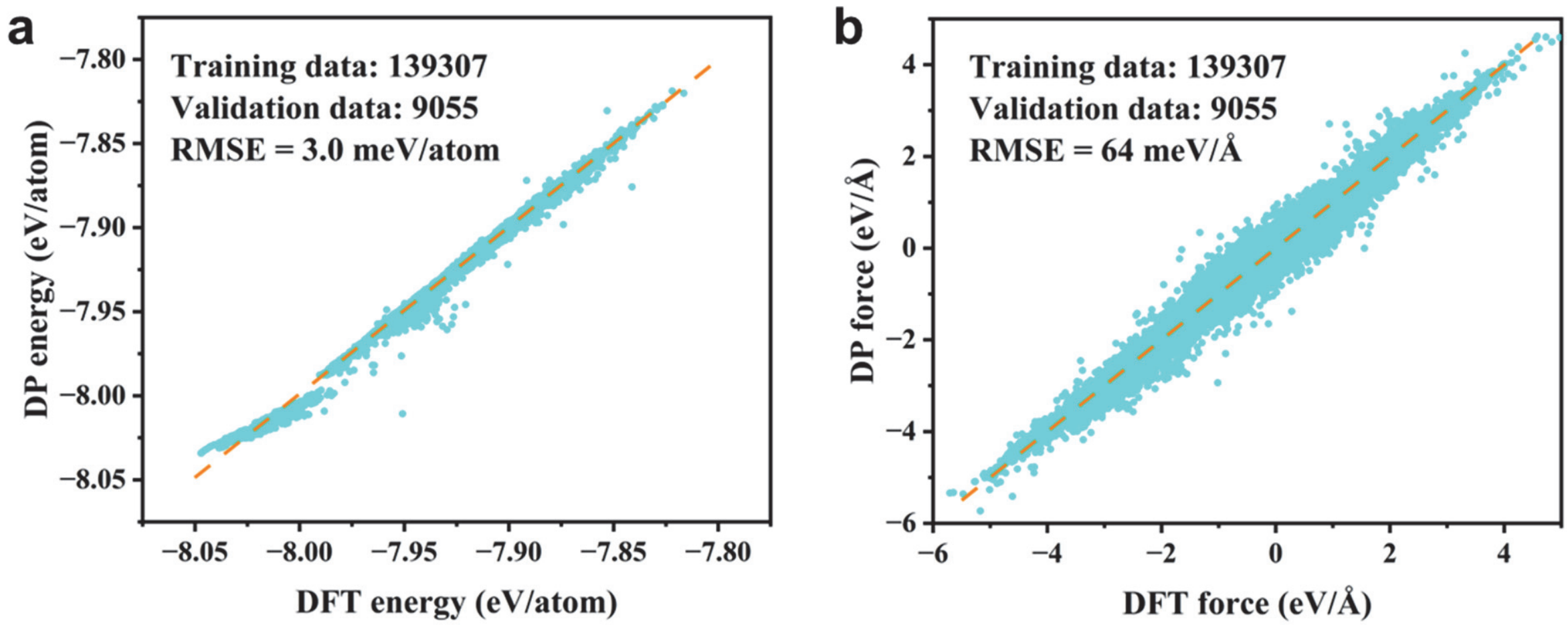

**Supplementary Fig. 3.** Training and validation results of the DPMD model for cerium oxide. Comparison of energies (**a**) and forces (**b**), respectively, predicted by DPMD with those calculated by DFT.

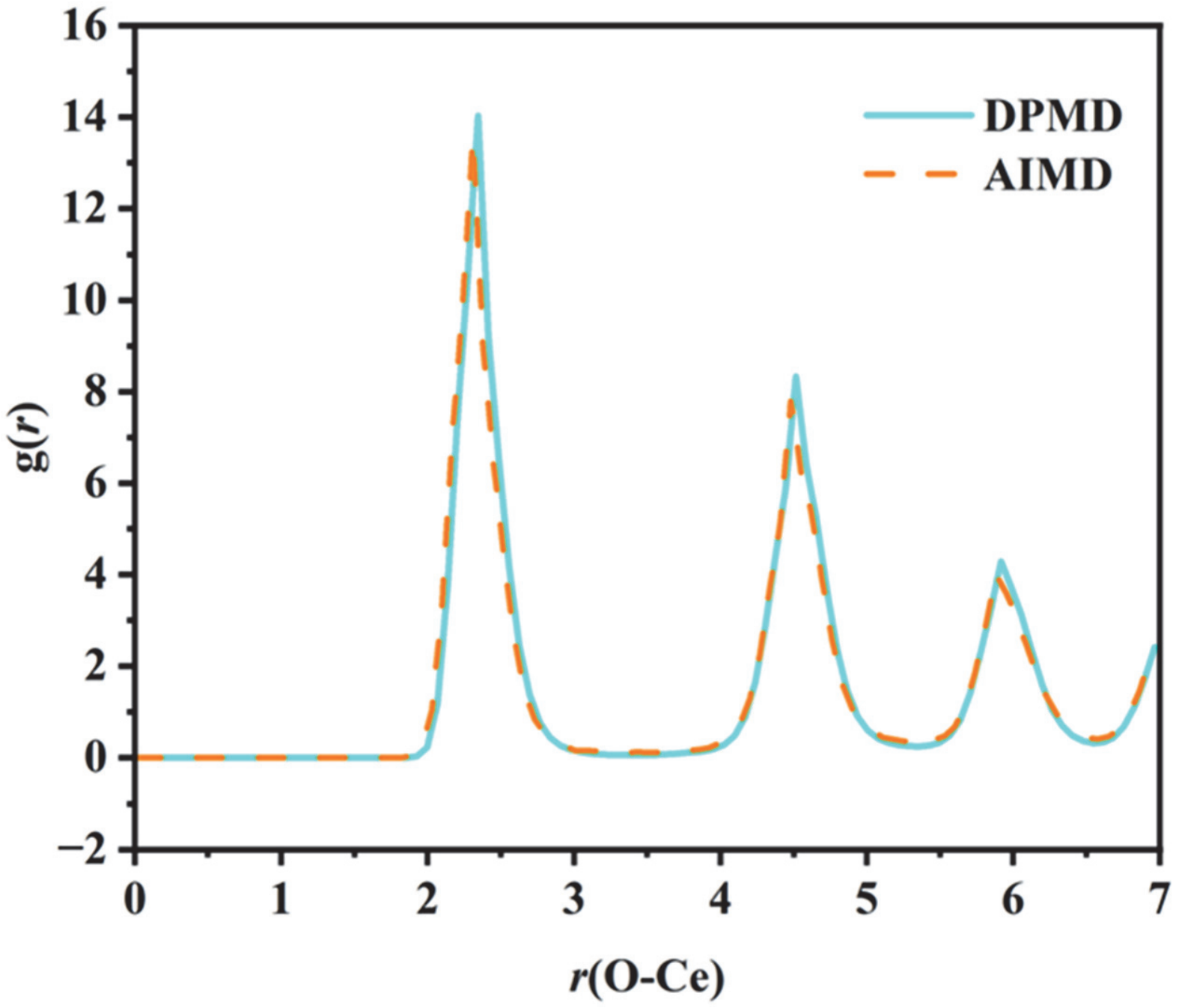

**Supplementary Fig. 4.** Radial distribution functions (RDF) for the O-Ce pair obtained from *ab initio* molecular dynamics and deep-potential molecular dynamics simulations at 900 K.

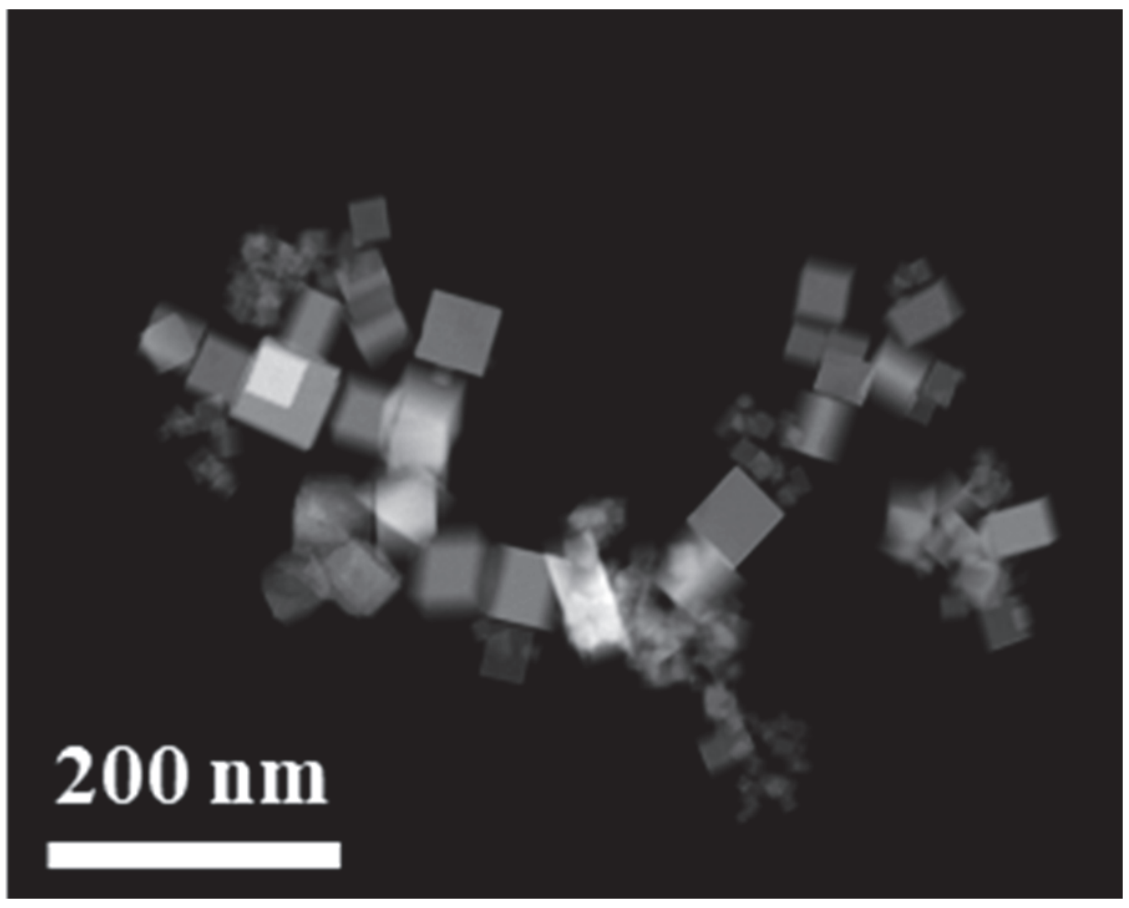

**Supplementary Fig. 5.** The low-magnification high-angle annular dark-field (HAADF) STEM image of $CeO_2$ nanocubes at 293 K in vacuum.

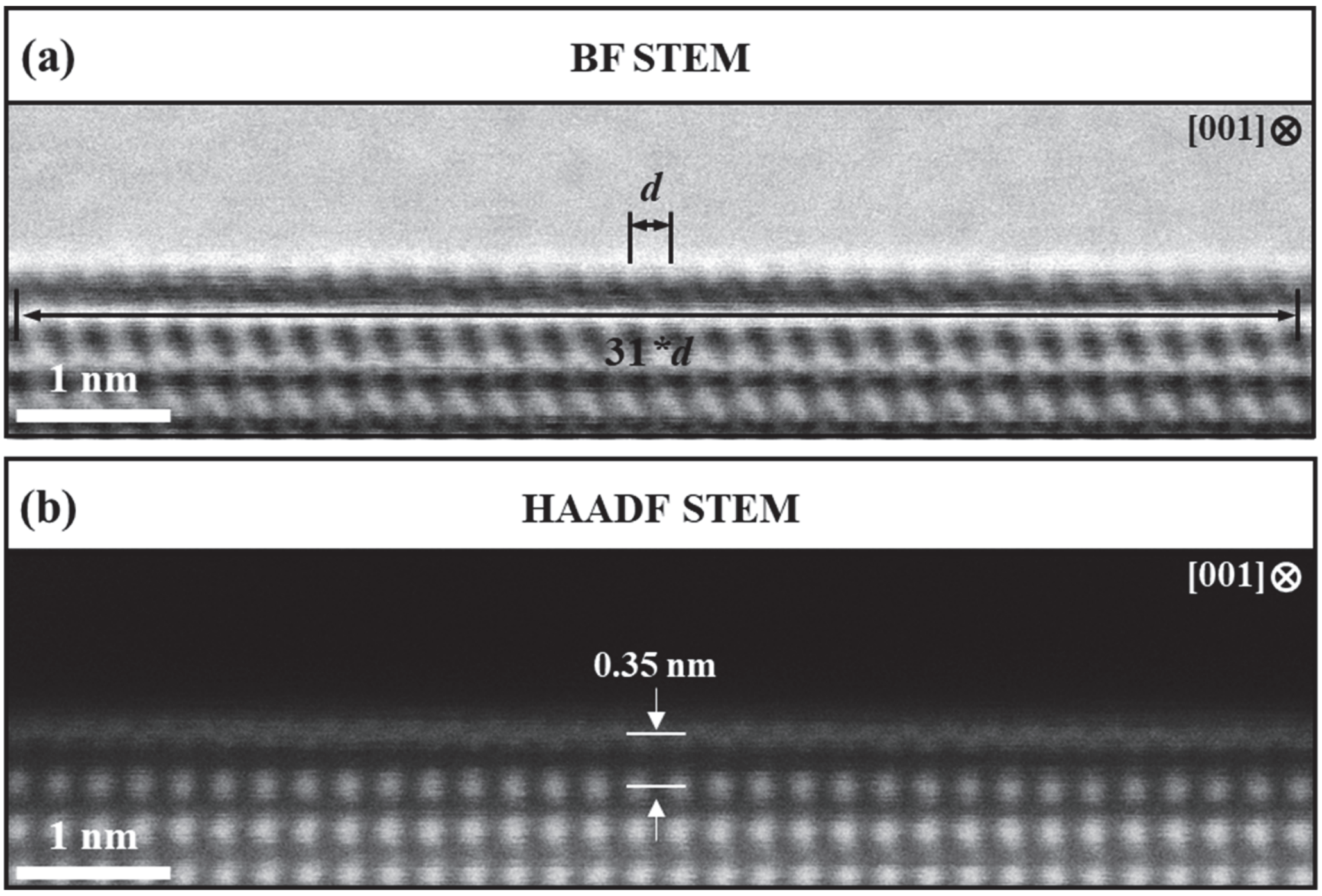

**Supplementary Fig. 6.** The in-situ atomic-resolution BF (Bright-field) (**a**) and HAADF (**b**) STEM images of long-range $CeO_2$ (100) surface along the [001] zone axis at 1173 K in vacuum (pressure: ~$10^{-5}$ Pa).